# Fingerprint Liveness Detection using Minutiae-Independent Dense Sampling of Local Patches

Riley Kiefer, Jacob Stevens, and Ashok Patel

**Abstract** Fingerprint recognition and matching is a common form of user authentication. While a fingerprint is unique to each individual, the authentication is vulnerable when an attacker can forge a copy of the fingerprint (spoof). To combat these spoofed fingerprints, spoof detection and liveness detection algorithms are currently being researched as countermeasures to this security vulnerability. This paper introduces a fingerprint anti-spoofing mechanism using machine learning. Specifically, the proposed algorithm builds upon existing methods of local patch-based fingerprint liveness detection by applying a dense and overlapping local patch sampling method and applying algorithm reduction principles to reduce algorithm complexity for embedded platforms. These minutiae-independent local patches are uniformly rotated according to the patch's intensity gradient. Once the preprocessing methods are completed, the patches are passed into a shallow Convolutional Neural Network (CNN) and an aggregate of the patch scores determines the fingerprint classification. This proposed method is tested on the public fingerprint liveness detection dataset: LivDet-2009 Biometrika, CrossMatch, and Identix as well as LivDet-2011 Biometrika, Digital, and Sagem. Intra-sensor models are generated, tested, and compared to other top algorithms created by researchers using the Average Classification Error (ACE) metric. A graphic user interface tailored to the proposed method is also presented to visualize the classifier results at the local level.

## 1 Introduction

Fingerprint authentication has become the norm for identification for many individuals, groups, and even businesses. Despite its popularity, fingerprint authentication has one major security threat: the ability to reproduce the original fingerprint and pose as an authorized user. Attackers can use a wide range of materials to hold the ridges of the artificially copied (spoofed) fingerprint. As a countermeasure to this fingerprint presentation attack, a system must examine the fingerprint and determine the fingerprint liveness classification before performing the fingerprint matching. Several researchers have proposed various systems to determine the liveness of a fingerprint: hardware approaches examining natural liveness phenomenon, software approaches using machine learning to understand the characteristics of live and fake fingerprint images, and a hybrid approach. This paper will examine fingerprint liveness detection using a software approach with machine learning as the model classifier. In addition to machine learning, several computer vision principles are applied as a preprocessing pipeline to transform a single fingerprint image into many local fingerprint patches. The classifier will train off these patches instead of the original fingerprint.

    This work is inspired by T. Chugh et al. [3]. Specifically, T. Chugh et al. introduced an effective patch-based fingerprint liveness detection algorithm. One of the key distinctions of this work compared to T. Chugh et al. [3] and the Spoof Buster algorithm is that the local patches in our algorithm are minutiae-independent while the Spoof Buster uses minutiae-dependent patches. This important distinction saves processing time in finding the minutiae, which could take a significant amount of time depending on the size of the image. Instead, generated patches in the preprocessing step are minute-independent and overlapping- which are extracted using a dense sampling method. In the effort of continuing to reduce the algorithm time complexity for a lightweight fingerprint liveness classifier, other modifications to the Spoof Buster [3] are necessary. The proposed model uses a shallow CNN for patch classification instead of a transfer-learned deep CNN, which will use less memory and disk space. Finally, T. Chugh et al. [3] uses an additional classifier (unspecified, but likely an Support Vector Machine- SVM), to determine the fingerprint classification from all extracted patches, while our work simply takes an aggregate of the patch scores for all patches in a fingerprint to determine the classification.

R. Kiefer, J. Stevens, A. Patel
Florida Polytechnic University
Lakeland, Florida, United States
{rkiefer, jstevens, apatel}@floridapoly.edu

The aggregate scoring simply takes a max sum to determine the classification, which is considerably less complex of an operation than an SVM classification. In summary, the objective of this work is to improve upon the efficiency of the preprocessing and testing by removing the patch dependence on the minutiae and by reducing model complexity while retaining a similar model accuracy.

This paper will review related works (Section 2), the proposed preprocessing algorithm for patch generation (Section 3), the machine learning algorithm and training details (Section 4), the intra-sensor model results when tested on the LivDet-2009 and LivDet-2011 fingerprint liveness detection database (Section 5) along with a graphical visualization of the local patch scoring results, and closing remarks with a conclusion and discussion on future work (Section 6). The algorithm is divided into the following stages: the preprocessing approach to extract dense sampled local patches, the local patch classifier, and the fingerprint classifier which aggregates the local classifications. The model results are also compared against other top algorithms presented in the survey of R. Kiefer et al. [1][2].

## 2  Related Works

Many recently published papers introduce novel approaches to fingerprint authentication, liveness detection, spoof recognition, etc. The surveys presented by R. Kiefer et al. [1][2] highlight the vast number of algorithms presented by researchers over the years, with varying levels of effectiveness. There are many different approaches to the recognition of a fingerprint. The task of fingerprint liveness classification often falls into the hands of a Convolutional Neural Network (CNN). The CNN is typically pipelined into some other classification model in a combinational approach. A very popular paper, "Fingerprint Spoof Buster: Use of Minutiae-Centered Patches" [3], employs a transfer-learned CNN on local minutiae-based and uniformly oriented local patches. With the local patches scored and passed into a second classifier, the fingerprint classification is generated. A similar technique of orientation adjustment was also seen in one algorithm where the initial data is segmented, rotated, then trained [4]. The papers that include orientation adjustment in their preprocessing pipeline typically had more effective algorithms results (whether it was global or local adjustments) compared to learning on the raw fingerprint data. The Spoof Buster algorithm [3] inspired the work of this paper because of the effectiveness of adjusted local patches as a preprocessing method.

Other popular approaches to liveness detection include pore detection systems as a recognition method [5]. By analyzing the pores of a fingerprint, the pores on a spoof are less defined or non-existing depending on the quality of the spoof, the sensor type, and if the subject was cooperating. While deep learning models are typically used, they are often expensive in terms of computational costs to train a model. Other work has been done on efficient model training and classification performance via a lightweight CNN [6]. The operational costs of software-based approaches typically run much cheaper than the costs of hardware approaches like thermal imaging or light cameras [7]. However, software-based approaches are typically more vulnerable to deep learning attacks [8] and cross-material/sensor attacks [9].

## 3  Proposed Preprocessing Approach

The proposed approach for the fingerprint liveness detection algorithm involves three steps: local patch extraction and preprocessing, local patch classification, and an aggregate fingerprint classification based on the local patch classifications. The following will discuss the preprocessing approach in detail.

### 3.1  Local Vector Orientation Field

The first step in the proposed algorithm's preprocessing stage is computing a matrix of the 2D local vector orientations based on pixel gradients for all grid cells in an image. A grid cell, τ, is defined as a σ x σ pixel region of the fingerprint image, where σ is equal to 12 pixels. This value was chosen arbitrarily as a base unit of measure. For illustrative purposes, the local vector orientation field would generate vectors for each grid cell of variable angle and magnitude as depicted with lines in Figure 1. These vectors typically align with the fingerprint ridges residing inside the grid cell, depending on the amount of noise and artifacts in the region.

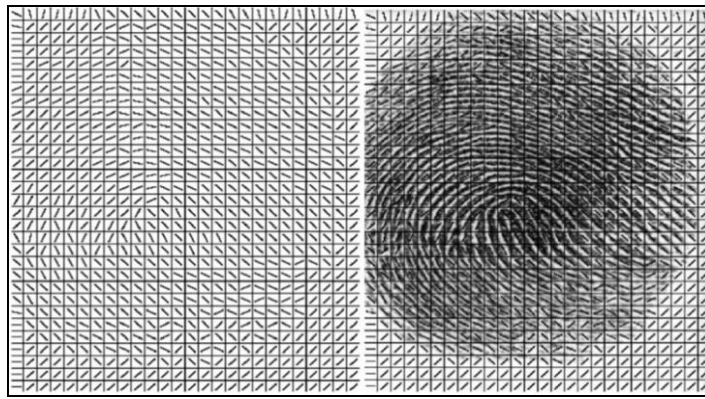

*Figure 1. Local Vector Orientation Field of Grid Cells (Left) and the Local Vector Orientation Field of Grid Cells Drawn on the Original Fingerprint Image (Right)*

The first step necessary to create a local vector orientation field of grid cells is to estimate the 1st order derivative of the pixel gradient using the central difference, δ, of pixel intensities for both the x and y axis of pixel ρ.

$$\delta_x\,(\rho[y][x]) = \rho[y][x+1] - \rho[y][x-1]$$
$$\delta_y\,(\rho[y][x]) = \rho[y+1][x] - \rho[y-1][x]$$

The absolute value of the difference is also calculated to obtain the directional magnitudes. By taking the magnitude, pixel intensity shifts at ridges are preserved rather than cancelled out. Now that the central difference is computed for a single pixel for both the x and y axis, the total central difference needs to be computed for the entire neighborhood of the grid cell, τ. This can be represented as a summation of the central difference for a given axis. Note that the edge pixels are not included in this summation because they may be on the border of the image.

$$\delta_{y,\tau}(\sigma,\rho[y][x])) = \sum_{j=2}^{\sigma-1}\sum_{i=2}^{\sigma-1}|\delta_y\,(\rho[y+j][x+i])|$$
$$\delta_{y,\tau}(\sigma,\rho[y][x])) = \sum_{j=2}^{\sigma-1}\sum_{i=2}^{\sigma-1}|\delta_y\,(\rho[y+j][x+i])|$$

After obtaining the grid cell magnitude in the x and y direction, the overall magnitude of the grid cell vector is computed.

$$magnitude(\tau) = \sqrt{\delta_{y,\tau}(\sigma,\rho[y][x]))^2 + \delta_{x,\tau}(\sigma,\rho[y][x]))^2}$$

Now, the magnitude is computed for every grid cell in the image and saved into a matrix for future access. Each grid cell's unit magnitude is stored at an index in the matrix for each axis x and y. Since unit magnitudes are only positive, it is also important to also note if the vector is upward or downward facing- this will be useful in determining the patch (a collection of grid cells) orientation.

$$unitMagnitudeMatrixX[j][i] = \frac{\delta_{x,\tau}(\sigma,\rho[y][x]))}{magnitude(\tau)}$$
$$unitMagnitudeMatrixY[j][i] = \frac{\delta_{y,\tau}(\sigma,\rho[y][x]))}{magnitude(\tau)}$$

## 3.2 Local Patch Extraction

A local patch, φ, is defined as a collection of grids cells τ. With the grid cell magnitudes computed for all grid cells and saved in a matrix (depicted visually in Figure 1), the local overlapping fingerprint patches can use the grid cell's magnitude to compute the local patch angle. Once the angle of a patch is found, the patch is rotated to 0 degrees (from the right), cropped to remove any background generated from rotating, and saved to the storage (Figure 2).

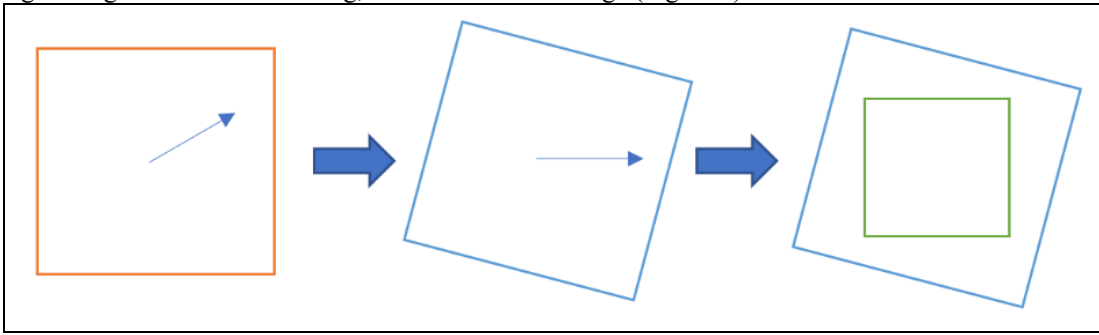

*Figure 2. The local patch generation process: patch angle computed (Left), patch rotation to a normalized angle (Middle), and patch cropping to a normalized size (Green) (Right)*

This process of patch orientation adjustment is conducted in a similar manner by T. Chugh et al. [3] and the Spoof Buster algorithm- the main difference is that patches generated in [3] are only generated from key-points in the fingerprint containing minutiae. Another important difference is that the patches generated in [3] typically are not overlapping unless some minutiae are densely packed together. In this work, all patches have some overlap with one another. Specifically, every new patch created is shifted over by the size of the grid cell, σ (as seen in Figure 3).

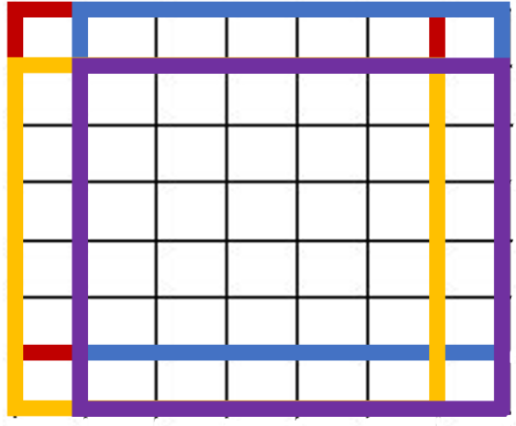

*Figure 3. An example of four overlapping patches (red, blue, yellow, and purple) created from a selection of grid cells (black boxes). In this simple example, the grid cell multiplier is 4 and the padding multiplier is 1, for a total patch size of 6x6.*

The first step in generating several local fingerprint patches from a fingerprint image is to first specify the size of the fingerprint patch. This patch size comprises of two parts: 1) a grid cell multiplier and 2) a padding multiplier. The grid cell multiplier is an integer multiplied by the grid size, $\sigma$, which represents the number grid cells wide and tall that the center of the patch contains. The padding multiplier is also an integer multiplied by the grid size, $\sigma$, which represents how many additional grid cells wide and tall that the patch contains along the edge. However, the padding grid cells do not contribute in determining the patch orientation. The only purpose that the padding serves is to provide additional local pixel data that can be used to fill in gaps that are lost due to the rotation and cropping. The patch components are depicted visually in the Figure 4 (left).

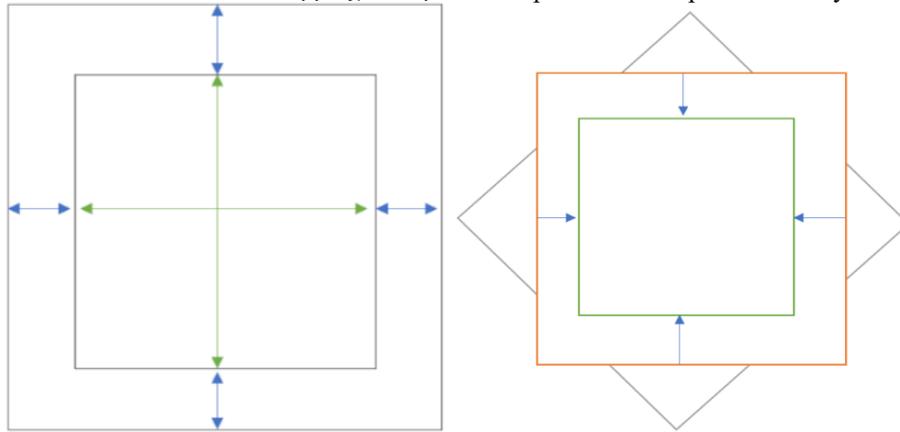

*Figure 4. Left: Local Fingerprint Patch Components: Green- Grid cells from the Grid Cell Multiplier. Blue- Grid cells from the Padding Multiplier. Right: Original Patch (Orange), Rotated Patch to the worst-case angle of 45 degrees (Grey), Cropped Patch (Green). The blue arrows represent half the crop value. Notice that the cropped patch is now contained within the rotated patch, eliminating any unwanted background.*

By precomputing a matrix of grid magnitudes, the generation of overlapping patches is much more efficient. Each new patch is shifted over by one grid cell ($\sigma$), which creates a lot of overlap (Figure 3) but will make each extracted patch unique. From the patch gradients, the patch orientation can be computed by summing all grid cell magnitudes in each patch. For patch, $\varphi$, the patch gradients can be represented as a summation of the grid cell unit vectors within the bounds of the patch (central grid cells only, not padding grid cells- see Figure 4).

$$\delta_{y,\varphi} = \sum_y \sum_x unitMagntiudeMatrixY[y][x]$$
$$\delta_{x,\varphi} = \sum_y \sum_x unitMagntiudeMatrixX[y][x]$$

With the total magnitude of the patch computed for both the x and y axis across all the patches' grid cells, the patch orientation angle is derived by trigonometry.

$$\theta = \tan^{-1}\left(\frac{\delta_{y,\varphi}}{\delta_{x,\varphi}}\right)$$

With the patch angle computed, the extracted patch (containing both the padding grid cells and the central grid cells) is rotated by $-\theta$ degrees so that all patches are uniformly rotated to 0 degrees (Figure 2). Since a rotation introduces a background fill color when the image dimensions increase, the patch is cropped to prevent any background color from being included in the result, despite the angle. To accommodate for any angle, the patch is assumed to be at the worst-case orientation, which 45 degrees. The amount of cropping on the x and y axis needed to remove all background fill color is calculated below.

$$crop = \left\lceil \sqrt{\frac{(\sigma \times patchMultiplier)^2}{8}} \right\rceil$$

With the crop amount calculated, the crop is split in half, so the cropping is distributed evenly between the left and right sides along the horizontal and top and bottom sides along the vertical. Figure 4 (right) depicts the cropping scenario for the worst case.

Now that the cropped patch has been generated, the last step is to check if the patch has enough pixel data before saving to the disk. The mean pixel intensity of the whole fingerprint image is compared against the mean pixel intensity of the patch. If the patch has a smaller mean (more data since a black pixel is represented as 0) than the whole fingerprint image's mean, the patch is saved, otherwise it is rejected. This is a simple way of rejecting patches with a large amount of whitespace and reducing training/testing computational complexity.

$$Patch\ Comparison = \begin{cases} if\ (mean_{image} < mean_{patch} - (mean_{patch} \times t)): save\ patch \\ else: \qquad\qquad\qquad\qquad\qquad\qquad\qquad\qquad reject\ patch \end{cases}$$
$$where\ 0 \leq t \leq 1\ is\ the\ noise\ factor$$

## 4   Classification Model

With the dataset preparation via fingerprint patch generation completed in the preprocessing stage, the fingerprint patches features need to be learned by a machine learning algorithm to classify the patches as real or fake. Once the fingerprint patches are classified, the fingerprint itself can be classified.

### 4.1   Local Patch Classification

Another important distinction of this work compared to T. Chugh et al. [3] and the Spoof Buster algorithm is that our classification model employs the use of shallow Convolutional Neural Network (CNN) instead of a transfer learning approach from the MobileNet-v1 architecture. The architecture is depicted in Figure 5.

With small patch sizes, it is easy to overfit a model with the millions of parameters trained in MobileNet. For this reason, a shallow CNN is more suited to learn the liveness characteristics of the small patches that are 82 pixels tall and wide. With a limited number of pixels, custom deep models with many convolutions is not possible, and very shallow models with few convolutions may not be effective enough to learn the distinctive liveness features. Another benefit of a shallow model is the efficiency for embedded implementations- with faster model load times and small disk usage. The chosen classification model has 4 convolutions, with each convolution doubling the number of filters from the previous convolution (64, 128, 256, 512) and each convolution increasing the dropout by 10% from the previous convolution (20%, 30%, 40%, 50%) to reduce model overfitting. Between the convolution and the dropout, Relu activation and batch normalization is applied. After the final dropout, the output is flattened and fully connected to 2 nodes for the binary classification problem. Finally, a soft-max layer is applied to extract the classification percentages for each class.

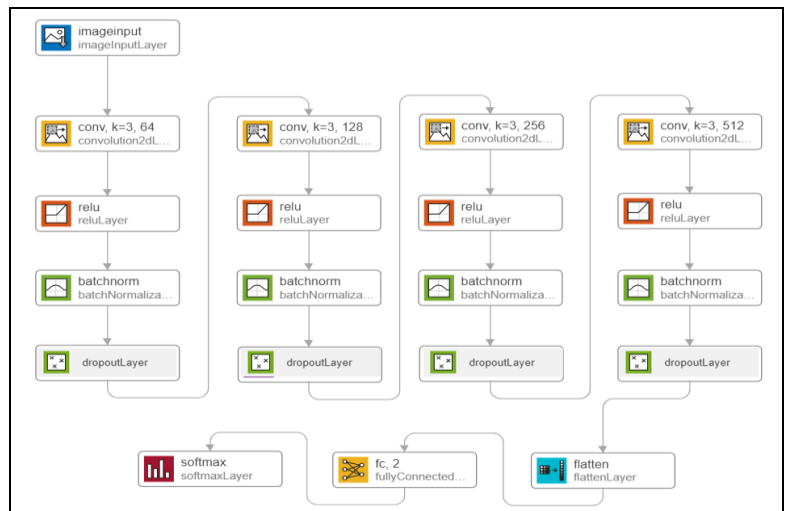

*Figure 5. Shallow CNN for local patch classification.*

The model is optimized using Adam, sparse categorical cross-entropy is used for the loss, and accuracy is the model metric. Model training uses batches of 32 patch images at a time. Several models were tested on different patch sizes using the same model, but larger patch sizes were trained for a longer number of epochs. Since the patches are all preprocessed to the same orientation angle, no transformative cropping or rotation data augmentation techniques are applied on the training set before model training, unlike the Spoof Buster [3] algorithm approach.

## 4.2 Aggregate Fingerprint Classification

The result of the output soft-max layer from the shallow CNN gives a liveness score percentage and a spoof score percentage for each patch. The sum of all live score percentages and spoof score percentages are computed for all generated patches for a given fingerprint image.

$$aggregate\ live\ score = \sum local\ live\ score$$
$$aggregate\ spoof\ score = \sum local\ spoof\ score$$

If a given fingerprint image has a larger global liveness score than global spoof score, the image is classified as real. Otherwise, the image is classified as fake. While some papers, like T. Chugh et al. [3] and the Spoof Buster algorithm employed the use of a variable learned threshold for distinguishing a real or fake fingerprint for slightly increased performance, the algorithm simply uses a max aggregate score for classification.

# 5 Experimental Results

## 5.1 Performance Metrics

The following metrics are used in determining the accuracy of our liveness detection model. The following shorthand is used: False Negative (FN), False Positive (FP), True Negative (TN), True Positive (TP).

### 5.1.1 FAR: False Acceptance Rate

$$FAR = \frac{FP}{FP + TP}$$

### 5.1.2 FRR: False Rejection Rate

$$FRR = \frac{FN}{FN + TN}$$

### 5.1.3 ACE: Average Classification Error

$$ACE = \frac{FAR + FRR}{2}$$

### 5.1.4 Accuracy

$$Accuracy = \frac{TP + TN}{TP + TN + FP + FN}$$

## 5.2 Parameters and Environment

All models used the following preprocessing parameters: a patch multiplier of 10, a padding multiplier of 2, and a grid size of 12. Based on some experimental testing with varying patch multiplier and padding values, model accuracy increased marginally with increases in the size of the patch multiplier ($\sigma$). For the LivDet-2009 Biometrika model, a patch multiplier of 6 and a pad of 1 resulted in a testing accuracy of 98.9%. When changed to a patch multiplier of 8 and a pad of 2, the testing accuracy increased to 99.2% accuracy. Finally, when changed to a patch multiplier of 10 and a pad of 2, testing accuracy increased to 99.4% accuracy. The noise factor, t, plays a role in determining how many patches are generated. Based on empirical testing with various noise factors, a threshold of 0.1 was found to produce on average, about 200 patches per fingerprint image.

All models trained for 30 epochs (roughly 20 hours depending on the dataset) on the generated patch training set generated for each of the scanners. Model training was conducted on a machine with the following specifications: i7-9700k Central Processing Unit (CPU), RTX 2080 Graphics Processing Unit (GPU), and a 2TB Hard Disk Drive (HDD) storage. The largest bottleneck was the HDD storage, which inhibited the rapid transfer of small patches to the RAM for testing. For more efficient model training, we recommend storing the generated patches on a Solid State Drive (SSD) or in memory.

## 5.3 Intra-Sensor Results for LivDet-2009

All models are trained and tested on the LivDet-2009 fingerprint liveness detection dataset [10] after the preprocessing patch generation step (Section 3) is applied on the raw fingerprint images. The LivDet-2009 dataset comprises of 3 scanners- Biometrika, CrossMatch, and Identix (Figure 6). The results of the local patch classification (Figure 7) and the aggregate fingerprint classification of all the local patch scores (Figure 8) is presented below.

| Dataset | Training | | Testing | | Spoof Material |
|---|---|---|---|---|---|
| | Live Samples | Spoof Samples | Live Samples | Spoof Samples | |
| *Biometrika* | 520 | 520 | 1473 | 1480 | Silicone |
| *CrossMatch* | 1000 | 1000 | 3000 | 3000 | Silicone, Gelatin, PlayDoh |
| *Identix* | 750 | 750 | 2250 | 2250 | Silicone, Gelatin, PlayDoh |

*Figure 6. LivDet-2009 dataset training and testing details for each scanner. Note that these counts represent the total number of fingerprint images, not the number of patches generated after preprocessing.*

| Scanner | FRR | FAR | ACE | Accuracy |
|---|---|---|---|---|
| *Biometrika* | 0.55% | 8.34% | 4.44% | 95.68% |
| *CrossMatch* | 8.40% | 5.69% | 7% | 92.78% |
| *Identix* | 14.42% | 0.21% | 7.31% | 90.95% |

*Figure 7. Intra-sensor local patch classification performance results for the LivDet-2009 dataset.*

| Scanner | FRR | FAR | ACE | Accuracy |
|---|---|---|---|---|
| *Biometrika* | 0% | 1.22% | 0.61% | 99.39% |
| *CrossMatch* | 0.16% | 0.90% | 0.53% | 99.34% |
| *Identix* | 4.48% | 0% | 2.24% | 97.66% |

*Figure 8. Intra-sensor fingerprint classification performance results after taking the aggregate score of all patches for a given fingerprint for the LivDet-2009 dataset.*

### 5.4 Intra-Sensor Results for LivDet-2011

All models are trained and tested on the LivDet-2011 fingerprint liveness detection dataset [16] after the preprocessing patch generation step (Section 3) is applied on the raw fingerprint images. The LivDet-2011 dataset comprises of 4 scanners, 3 of which were tested on (note: Italdata was not tested)- Biometrika, Digital, and Sagem (Figure 9). The results of the local patch classification (Figure 10) and the aggregate fingerprint classification of all the local patch scores (Figure 11) is presented below.

| Dataset | Training | | Testing | | Spoof Material |
|---|---|---|---|---|---|
| | Live Samples | Spoof Samples | Live Samples | Spoof Samples | |
| *Biometrika* | 1000 | 1000 | 1000 | 1000 | EcoFlex, Gelatin, Latex, Silgum, Wood Glue |
| *Digital* | 1004 | 1000 | 1000 | 1000 | Gelatin, Latex, Playdoh, Silicone, Wood Glue |
| *Italdata* | 1000 | 1000 | 1000 | 1000 | EcoFlex, Gelatin, Latex, Silgum, Wood Glue |
| *Sagem* | 1008 | 1007 | 1000 | 1036 | Gelatin, Latex, Playdoh, Silicone, Wood Glue |

*Figure 9. LivDet-2011 dataset training and testing details for each scanner. Note that these counts represent the total number of fingerprint images, not the number of patches generated after preprocessing.*

| Scanner | FRR | FAR | ACE | Accuracy |
|---|---|---|---|---|
| *Biometrika* | 7.30% | 3.11% | 5.21% | 95.06% |
| *Digital* | 5.91% | 8.32% | 7.12% | 92.81% |
| *Sagem* | 9.71% | 2.53% | 6.12% | 94.08% |

*Figure 10. Intra-sensor local patch classification performance results for the LivDet-2011 dataset.*

| Scanner | FRR | FAR | ACE | Accuracy |
|---|---|---|---|---|
| *Biometrika* | 2.30% | 4.40% | 3.35% | 96.04% |
| *Digital* | 2.30% | 1.00% | 1.65% | 98.35% |
| *Sagem* | 1.06% | 4.50% | 2.78% | 97.08% |

*Figure 11. Intra-sensor fingerprint classification performance results after taking the aggregate score of all patches for a given fingerprint for the LivDet-2011 dataset.*

## 5.5 Comparison to Other Algorithm Results

The aggregate fingerprint classification ACE metric for each scanner (Figure 12) is compared against the top published algorithms submitted by researchers. Using the survey on fingerprint presentation attacks provided by R. Kiefer et al. [1][2], our proposed model ranking for each scanner is determined.

| Rank | Reference | Algorithm Name | ACE (%) |
|---|---|---|---|
| 1 | 11 | DCNN and SVM, RBF Kernel | 0 |
| 2 | 12 | Local Uniform Comparison Image Descriptor (LUCID) | 0.14 |
| 3 | 11 | DCNN and SVM, Polynomial Kernel Order 2 | 0.38 |
| 4 | 11 | DCNN and SVM, Polynomial Kernel Order 3 | 0.57 |
| **5** | **-** | **PROPOSED MODEL** | **0.61** |

*Figure 12. Top 5 intra-sensor models ranked by the ACE metric from the results presented in the survey [1][2] for LivDet-2009 Biometrika. Our proposed model ranks 5$^{th}$ place.*

| Rank | Reference | Algorithm Name | ACE (%) |
|---|---|---|---|
| **1** | **-** | **PROPOSED MODEL** | **0.53** |
| 2 | 13 | CNN-VGG-227 | 0.6 |
| 3 | 14 | MvDA: G5, SID RICLBP LCPD DSIFT | 1 |
| 4 | 13 | CNN-Alexnet-224x224 | 1.1 |
| 5 | 15 | Deep Triplet Embedding (Tnet) | 1.57 |

*Figure 13. Top 5 intra-sensor models ranked by the ACE metric from the results presented in the survey [1][2] for LivDet-2009 CrossMatch. Our proposed model ranks 1$^{st}$ place.*

The proposed algorithm ranked 5th place with an ACE of 0.61% on the LivDet-2009 Biometrika dataset (Figure 12), 1st place with an ACE of 0.53% on the LivDet-2009 CrossMatch dataset (Figure 13), and the LivDet-2009 Identix model did not get a top placement with an ACE of 2.24%. Our model's average LivDet-2009 ACE is 1.12%, which places our algorithm as the 6th best overall algorithm on LivDet-2009 according to the survey results [1][2]. On the LivDet-2011 database, none of our models received a top placement.

Our proposed model is now compared to the Spoof Buster algorithm [3] in terms of ACE comparison and processing time. LivDet-2011 is the only dataset that was tested on both the proposed model and the Spoof Buster model. The Biometrika and Sagem ACE only differed by several percentage points when compared to the Spoof Buster. On the other hand, the Ditigtal ACE was very similar to the Spoof Buster performance. In terms of the classification time, the Spoof Buster model takes 100ms to classify a given image, and our proposed model takes 2000ms. Note that our solution was implemented in Python and was not optimized during our testing. Additionally, our models were tested in two different implementation environments- so it is difficult to compare our model efficiency when running on two different machines. Finally, this classification time reflects a fingerprint with 200 patches- compared to the 48 patches generated by Spoof Buster. In the conclusion, future work will detail how this can be optimized by random sampling these patches instead of using all generated patches for classification. Overall, despite the changes in patch extraction methods and model classifier changes, accuracy levels are relatively similar.

|  | Biometrika ACE (%) | Digital ACE (%) | Italdata ACE (%) | Sagem ACE (%) | Average Classification Time |
|---|---|---|---|---|---|
| **Proposed** | 3.35% | 1.65% | - | 2.78% | 2000ms |
| **Spoof Buster** | 1.24% | 1.61% | 2.45% | 1.39% | 100ms |

## 5.5 Comparison to Other Algorithm Results

Inter-sensor performance is also an important measure in determining an algorithm's performance. Inter-sensor performance measures how robust an algorithm is when tested on a dataset created with a different fingerprint scanner. Differences like image compression type, image size, and image resolution may impact the transferability of a model to another dataset. Additionally, testing on novel or untrained spoof materials may fool the classifier. While the intra-sensor models tested on the LivDet-2009 dataset performed very well according to the survey [1][2] results compiling the other published research, our algorithm's inter-sensor results leave much room for improvement.

Using the same model trained on the LivDet-2009 Biometrika patches dataset (containing silicone spoof materials), this model is tested on a different dataset- the LivDet-2009 CrossMatch and Identix patches dataset (containing gelatin, play-doh, and silicone spoof materials). The results are displayed in Figure 15. Note that aggregate testing was not performed here due to poor local accuracy. The aggregation strategy only works well if the local accuracy is acceptable. As evident in the data provided in the table, the LivDet-2009 Biometrika model does not transfer well to other datasets. The high False Acceptance Rate (FAR) is particularly alarming because this metric represents how often a fake fingerprint is accepted as a live fingerprint. While this test could not isolate whether the different image sensor or the novel materials played a bigger role in the significant drop in local patch accuracy, it is clear that this algorithm will need to be refined in future work.

| Scanner | FRR | FAR | ACE | Local Patch Accuracy |
|---|---|---|---|---|
| *CrossMatch* | 21.47% | 77.92% | 49.69% | 53.72% |
| *Identix* | 8.99% | 95.1% | 52.04% | 47.23% |

*Figure 15. Inter-sensor performance results for the LivDet-2009 Biometrika Model tested on the CrossMatch and Identix testing patch dataset.*

### 5.6 Local Patch Classifier Visualization

By following a similar approach to the T. Chugh et al. [3] and the Spoof Buster algorithm's graphical user interface, we developed a similar graphical patch scoring method tailored to our minutiae-independent dense sampling approach. The results are viewable in Figure 16. The colored grids represent the centers of the local patches. Red indicates that the local patch model classified the patch as fake. Green indicates that the local patch model classified the patch as live. If this algorithm was ever deployed, this visualizer can give access control personnel insight into identifying fingerprint spoofing at the local level. Specifically, with a patch-based algorithm, partial-spoof fingerprints can be identified and localized.

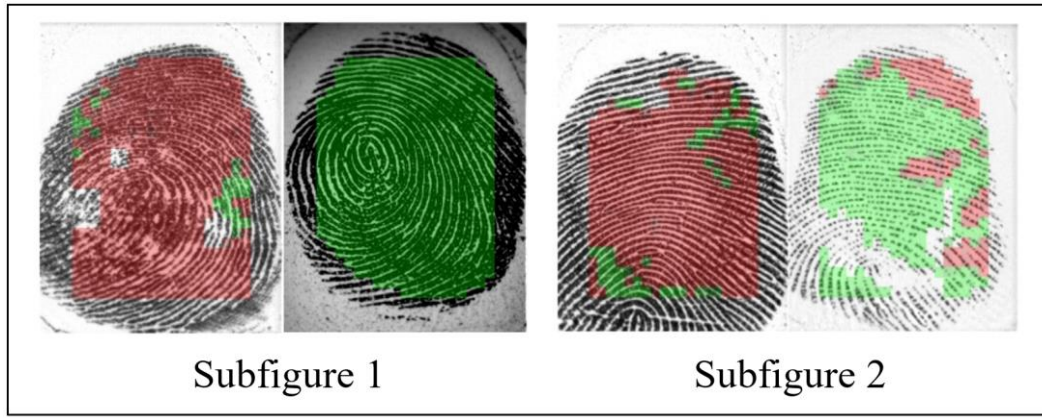

*Figure 16. Subfigure 1: Correctly classified spoof (left) and live (right) Biometrika fingerprint images. Subfigure2: Incorrectly classified false negative (left) and false positive (right) Biometrika fingerprint images.*

# 6 Conclusion

The objective of this work is to modify some of the major implementation choices of T. Chugh et al. [3] as a means of reducing algorithm complexity and testing speed without sacrificing accuracy. The proposed algorithm has several advantages and disadvantages.The first major difference is that patches are extracted in a dense and overlapping sampling approach, independent of the minutiae present in the fingerprint. By omitting the minutiae-detector, processing time is saved in iterating through the original fingerprint image for a set of patches centered around the minutiae. With the minutiae-detector, there are roughly 50 patches generated per fingerprint image, compared to roughly 200 patches generated per finger (depending on the image size and hyperparameters) using a dense sampling approach used in this research. While this large number of patches could be reduced considerably via the noise factor, t, our algorithm spends considerably more time training and classifying the local fingerprint patches due to the quantity alone. Future work could investigate the effects of random sampling of the patches generated from a given test fingerprint to see how well a subset of patches performs compared to all patches generated from a test fingerprint. This would reduce the testing time significantly, depending on the number of patches generated and tested.

Another important distinction is that our algorithm uses a shallow convolutional neural network, compared to the transfer learning approach adopted by T. Chugh et al. as the local fingerprint classifier. This type of model was chosen as a lightweight alternative for faster model load times and less disk usage with embedded systems in mind. It is also worth noting that the hyperparameter, $\sigma$ (grid size), also plays a role in determining the patch density. A larger $\sigma$ would produce less patches (but much larger patches) and a smaller $\sigma$ would produce more patches (but much smaller patches). Future work could further investigate how different combinations of grid size, patch multiplier, and padding multiplier can preserve model accuracy while reducing computational cost. Finally, the global fingerprint classifier of the Spoof Buster algorithm [3] uses a learned threshold, likely via a Support Vector Machine that trains on the aggregate local patch scores. Our algorithm simply employs a maximum of the aggregate live and spoof score as the classification, which also reduces algorithm complexity marginally. In a speed comparison test, we anticipated our model to run faster based on theoretical assumptions of reduced preprocessing steps and lighter classifiers. However, the Spoof Buster was much faster than our implementation, which is likely attributed to our quantity of patches generated and unoptimized code.

The dense sampling approach with minutiae-independent patches proves successful for intra-sensor fingerprint liveness classification. Our LivDet-2009 Biometrika model placed 5th best overall algorithm (3rd best unique algorithm) and our LivDet-2009 CrossMatch placed 1st overall based on the results published by R. Kiefer et al. [1][2]. Even though the other datasets did not receive top placement, they all had competitive accuracy levels. The LivDet-2011 results were nearly on par with the Spoof Buster algorithm [3], despite the model changes. While the local patch classifier across all scanner models did a satisfactory job in classifying all fingerprint patches at around 95% accuracy, the aggregate classifier overcomes the discrepancies at the local level by taking a summation of the local results to determine the fingerprint classification. This aggregate scoring of local classifications is much more effective in determining the fingerprint classification, with roughly a 99% accuracy. However, the algorithm failed to preserve its accuracy in the cross-sensor/cross-material test. When the model trained on the LivDet-2009 Biometrika dataset is tested on the LivDet-2009 Identix and CrossMatch dataset, the local accuracy nearly halved with 53% and 47% respectively. Cross-sensor and cross-material presentation attacks is an open-set problem in the field of fingerprint biometrics. While a defender has control of the sensor hardware, the attacker has the advantage by using novel fingerprint spoof materials that can fool a machine learning model trained on a specific subset of materials. While the intra-sensor (Biometrika) and intra-material (live and silicone) accuracy is very good, the inter-sensor and inter-material robustness will be addressed in future work to successfully thwart novel presentation attacks. Additionally, future work will test the same model on varying datasets to see if the accuracy levels are relatively stable when trained with a different scanner or with different spoof materials.

# 7 Acknowledgement

We would like to thank LivDet [10][16] for supplying the fingerprint dataset.